\documentclass[%
 reprint,
groupedaddress,
 amsmath,amssymb,
 aps,
prmaterials,
]{revtex4-1}
\usepackage{CJK}
\usepackage{enumerate}
\usepackage{graphicx}
\usepackage{dcolumn}
\usepackage{bm}
\usepackage{graphicx}
\usepackage{bbold} 
\usepackage{graphicx}
\usepackage{dcolumn}
\usepackage[left=3cm,top=2.5cm,right=3cm,bottom=2.5cm]{geometry}
\usepackage{array}
\usepackage{epsfig}
\usepackage{wrapfig}
\usepackage{color}
\usepackage{soul}
\usepackage{braket}
\usepackage{verbatim}
\usepackage{mathtools}
\usepackage{mathrsfs}
\usepackage{amsfonts}
\usepackage[english]{babel}
\usepackage{textcomp}
\usepackage{float}
\usepackage{comment}
\usepackage{url}
\usepackage{multirow}

\excludecomment{toexclude}
\usepackage{changepage} 

\begin{document}

\begin{CJK*}{UTF8}{}
\CJKfamily{gbsn}

\title{Density-Functional Theory and Triply-Periodic Minimal Surfaces}

\author{Mengdi Yin (尹梦迪), Jing Zhang (张\CJKfamily{bsmi}璟), Dimitri D. Vvedensky}
\affiliation{ The Blackett Laboratory, Imperial College London, London SW7 2AZ, United Kingdom}

\date{\today}
\begin{abstract}
Several authors have suggested that the surfaces of vanishing potential generated by the electrostatic fields from a distribution of point charges resemble triply periodic minimal surfaces (TPMS) corresponding to the positions of the point charges. We provide a theoretical basis for this phenomenological comparison by starting with the Boltzmann equation to show that the surface corresponding to zero charge density is a minimal surface. We then use density-functional calculations for elemental materials that differ electronically and structurally, Na, Cu, and Al, to show that surfaces of vanishing charge density converge to the corresponding TPMS.						

.
\end{abstract}


\maketitle
\end{CJK*}

Surfaces that minimize their area for given boundary conditions are called minimal surfaces. Historically, the concept of a minimal surface was proposed for the formation of a soap film, requiring the minimization of the surface free energy, which is proportional to the surface area \cite{hyde96,fomenko91}. 

The local shape near a point of a smooth three-dimensional surface is determined by the principal curvatures $k_1$ and $k_2$ of the maximum and minimum bending near that point. The local geometry of the surface is characterized by the mean curvature $M={1\over2}(k_1+k_2)$ and the Gaussian curvature $K=k_1k_2$  \cite{hyde96,carmo16}. Minimal surfaces with given boundaries have minimal free energy, such that the minimal surface area is ensured by the vanishing of $M$ \cite{hyde96,carmo16}.  The catenoid (the surface of revolution of a catenary) is a standard example of a minimal surface. 

\begin{figure}[t]
\centering
\includegraphics[width=0.9\linewidth]{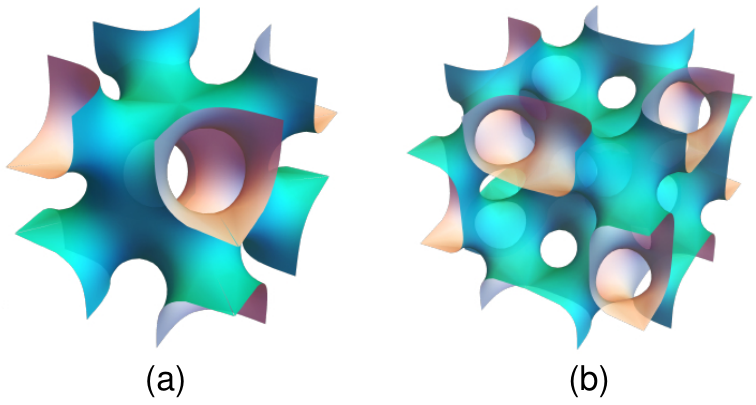}
\caption{Periodic units that repeat along three independent directions to form a TPMS: (a) the I-WP surface, for body-centered lattices, and (b) the F-RD surface, for face-centered lattices.}
\label{fig1}
 \end{figure}

Minimal surfaces with repeating units in three independent directions are called triply periodic minimal surfaces (TPMS) \cite{schoen70,schwarz90}. Examples of such surfaces are shown in Fig.~\ref{fig1}.  TPMS have attracted considerable interest in engineering and the physical and life sciences in areas as diverse as biology, chemistry, and materials science, including energy conversion, and the scaffold design of human tissue \cite{andersson88,torquato04,kapfer11,han18,alketan19}. 

Several authors \cite{schnering87,schnering91,mackay93,gandy01,gandy02} have suggested that TPMS are approximated by nodal surfaces of Fourier series, or as the zero equipotential surfaces (ZEP) of distributions of point charges. The main advantage of the Fourier series method is that TPMS are approximated by simple analytic combinations of trigonometric functions \cite{schnering91}.  

In this Letter, we take a more fundamental approach by using the Boltzmann equation to provide a semi-classical proof that the surface with zero charge density is a minimal surface.  We then use calculations based on density-functional theory (DFT) to determine constant charge density surfaces for several crystalline materials. We show that the surfaces of vanishing charge density do indeed correspond to the TPMS of the structure for the material. 

We begin with the semiclassical transport theory based on the Boltzmann equation \cite{landau81},
\begin{equation}
{\partial f\over\partial t}+{q\bm{F}\over \hbar}\cdot\bm{\nabla}_{\bm{k}}f+\bm{v}\cdot\bm{\nabla}_{\bm{x}}f=Q(f),
\end{equation}
where $f(\bm{x},\bm{k},t)\,d\bm{x}\,d\bm{k}$ is the probability of finding a particle in the volume $d\bm{x}\,d\bm{k}$ centered at $(\bm{x},\bm{k})$ at time $t$, $q$ is the charge on the particle, $\bm{F}$ is an external field, and $Q(f)$ is the collision integral.  In the low-density approximation for $Q$ \cite{jungel09}, the Boltzmann equation reduces to the continuity equation for the charge density $\rho$,
\begin{equation}
{\partial\rho\over\partial t}+\bm{\nabla}\cdot\bm{j}=0,
\label{eq2}
\end{equation}
where the current density $\bm{j}$ is the sum of drift $\bm{j}_F$ and diffusion $\bm{j}_D$ current densities:
\begin{equation}
\bm{j}_{\bm{F}}=q \mu \rho \bm{F}\, ,\qquad \bm{j}_D=qD\bm{\nabla}\rho.
\label{eq3}
\end{equation}  
Here, $\mu $ is the electron mobility and the Einstein relation \cite{reif65} is used for the diffusion coefficient $D$, which is assumed to be spatially homogeneous. In the stationary limit, (\ref{eq2}) is replaced by
\begin{equation}
\bm{\nabla}\cdot\bm{j}=\bm{\nabla}\cdot(\bm{j}_{\bm{F}}+\bm{j}_D)=0,
\label{eq3a}
\end{equation}
with the divergences
\begin{align}
\label{eq4}
\bm{\nabla}\cdot\bm{j}_F&=q\mu\bm{\nabla}\rho\cdot\bm{F}+q\mu\rho\bm{\nabla}\cdot\bm{F} ,\\
\bm{\nabla}\cdot\bm{j}_D&=qD\bm{\nabla}^2\rho.
\label{eq5}
\end{align}

\begin{figure}[t!]
\centering
\includegraphics[width=0.9\linewidth]{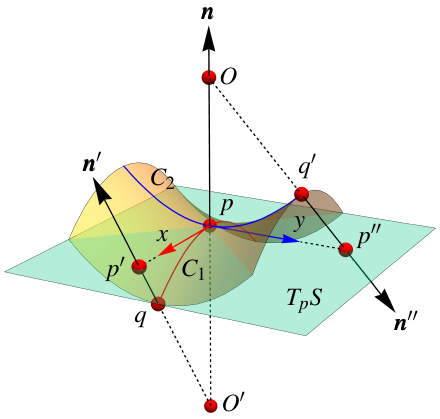}
\caption{Region of a surface (shown in yellow) near a point $p$, curves $C_1$ and $C_2$ on the surface along the $x$- and $y$-axes, respectively, and the tangent plane $T_pS$ at $p$ to the surface.}
\label{fig2}
\end{figure}

Consider a small section near a point $p$ on a surface of constant charge density (Fig.~\ref{fig2}). At $p$,
\begin{equation}
\partial_x\rho\big|_{p}=\partial_y\rho\big|_{p}=0,
\label{eq6}
\end{equation}
where subscript notation is used for partial derivatives, e.g., $\partial_x\equiv\partial/\partial x$ and $\partial_y\equiv\partial/\partial y$.  Although Fig.~\ref{fig2} displays the surface near $p$ as a saddle point, our proof makes no assumption about this surface other than corresponding to a constant charge density. We will prove that the surface with zero charge density is a minimal surface. In particular, (\ref{eq6}) is valid for any constant charge density surface.

The divergence theorem for $\bm{j}$ is,
\begin{equation}
\iint_S\bm{j}\cdot\bm{dS}=\iiint_V\bm{\nabla}\cdot\bm{j}\,dV=0,
\label{eq10}
\end{equation}
where $\bm{dS}$ is the outward normal of surface enclosing the volume $V$, and (\ref{eq3a}) is used to evaluate the right-hand side.  The left-hand side of (\ref{eq10}) is evaluated as an integral over the surface of the pill box centered at $p$. Because the tangential components of the current densities are integrated over an area whose width approaches zero, the integral over the sides of the pill box vanish.
The normal components of the drift and diffusion current densities are integrated over the top and bottom of the pill box. If these integrals are nonzero, charge either accumulates or is depleted within the pill box.  Hence, the invariance of the total charge of the surface enclosed by the pill box gives
\begin{equation}
\partial_{\bm{n}}j_{\bm{n}}=0,
\end{equation}
where $\partial_{\bm{n}}$ is the partial derivative with respect to the unit normal $\bm{n}$ to the surface at $p$ (Fig.~\ref{fig2}).  Thus, according to (\ref{eq3a}),
\begin{equation}
\bm{\nabla}\cdot\bm{j}=\partial_x\bm{j}_x+\partial_y\bm{j}_y,
\end{equation}
in which
\begin{equation}
\partial_\alpha\bm{j}_\alpha=\partial_\alpha\bm{j}_{\bm{F},\alpha}+\partial_\alpha\bm{j}_{D,\alpha},
\end{equation}
for $\alpha=x,y$. 
By using (\ref{eq4}), (\ref{eq5}), and (\ref{eq6}),  we have
\begin{align} 
\label{eq8}
\partial_\alpha\bm{j}_{\bm{F},\alpha}&=q\mu\rho\partial_\alpha\bm{F}_\alpha,\\
\partial_\alpha\bm{j}_{D,\alpha}&=qD\partial_{\alpha\alpha}\rho\, .
\label{eq9}
\end{align}
As $\rho\to0$, then, according to (\ref{eq8}) $\partial_\alpha\bm{j}_\alpha\to0$, so we are left with (\ref{eq9}), which yields
\begin{equation}
\partial_{xx}\rho+\partial_{yy}\rho=0\, .
\label{eq10a}
\end{equation}

\begin{figure}[b!]
\centering
\includegraphics[width=0.9\linewidth]{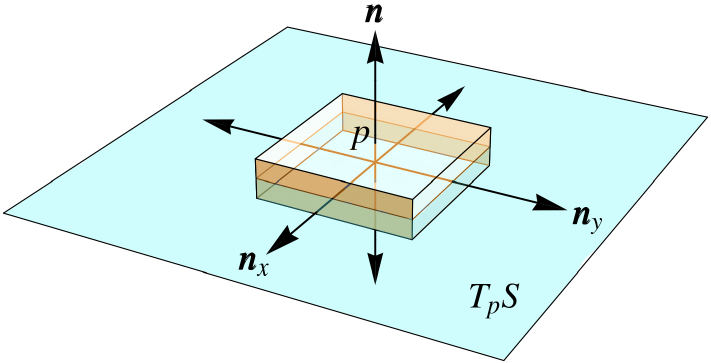}
\caption{Thin pill box on the tangent surface $T_pS$ centered at $p$ (Fig.~\ref{fig2}) used to calculate the divergence of the drift and diffusion current densities. The outward normal of each section are indicated.}
\label{fig3}
 \end{figure}

\begin{figure}[t!]
\centering
\includegraphics[width=\linewidth]{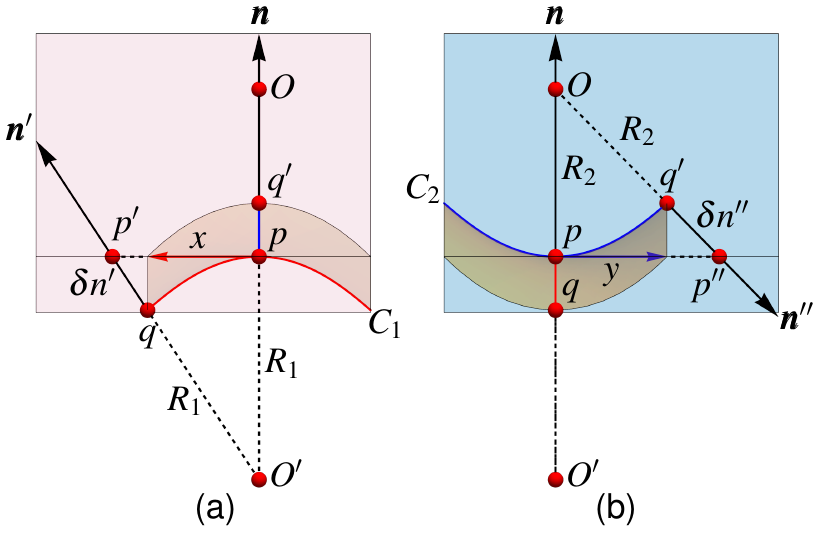}
\caption{Sections near the saddle point in Fig.~\ref{fig2} parallel to (a) the $x$-axis, and (b) the $y$-axis, with the same labelling. $R_1$ and $R_2$ are the radii of circles $C_1$ and $C_2$, respectively, that best approximate the surface along the $x$- and $y$-axes.  The curvature of the surface along the axes is the reciprocal of the radii.}
\label{fig4}
\end{figure}

The effects of curvature on diffusion are determined by considering the charge density at points $p^\prime$ and $p^{\prime\prime}$, which lie on the tangent plane of $p$, and points $q$ and $q^\prime$, which lie on the surface of constant charge density [Fig.~\ref{fig2} and Fig.~\ref{fig4}(a)].  By taking Taylor expansions of $p^\prime$ around $p$ ($\delta x=0$) and $q$ ($\delta n^\prime=0$), we have:
\begin{align}
\label{eq12}
\rho(p^\prime) &\approx\rho(p)+\partial_x\rho\big|_p\delta x+\textstyle{1\over2}\partial_{xx}\rho\big|_p(\delta x)^2\\
&\approx\rho(q)+\partial_{n^\prime}\rho\big|_q\delta n^\prime\, ,
\label{eq13}
\end{align}
where, because of (\ref{eq6}), the first non-vanishing term in (\ref{eq12}) is of order $(\delta x)^2$. From Pythagoras' theorem,  we have [Fig.~\ref{fig4}(a)]:
\begin{equation}
(\delta n^\prime+R_1)^2=R_1^2+\delta x^2,
\end{equation}
which yields, to leading order in $\delta x$,
\begin{equation}
\delta n^\prime\approx\frac{\delta x^2}{2R_1}.
\end{equation}
Given that $\rho(p)=\rho(q)$ and $\partial_x\rho\big|_p=0$, we have, from (\ref{eq12}) and (\ref{eq13}),
\begin{equation}
\frac{1}{R_1}\partial_n^\prime\rho\big|_q=\partial_{xx}\rho\big|_p.
\label{eq15}
\end{equation}

Similarly, by taking Taylor expansions of $p^{\prime\prime}$ about $p$ ($\delta y=0$) and $q^\prime$ ($\delta n^{\prime\prime}=0$) [Fig.~\ref{fig2} and Fig.~\ref{fig4}(b)], we obtain
\begin{equation}
\frac{1}{R_2}\partial_{n^{\prime\prime}}\rho\big|_{q^\prime}=\partial_{yy}\rho\big|_p.
\label{eq16}
\end{equation}
As $q$ and $q^{\prime}$ approach $p$,
\begin{equation}
\partial_n \rho\big|_p\approx\partial_{n^\prime}\rho\big|_q\approx\partial_{n^{\prime\prime}}\rho\big|_{q^\prime}.
\label{eq17}
\end{equation}
Using (\ref{eq15}), (\ref{eq16}), and (\ref{eq17}), the diffusion terms in (\ref{eq10a}) can be written as
\begin{equation}
(\partial_{xx}\rho+\partial_{yy}\rho)\big|_p=-\bigg( \frac{1}{R_1}+\frac{1}{R_2}\bigg)\partial_n\rho\big|_p.
\label{18}
\end{equation}
According to (\ref{eq10a}), the left-hand side of this equation must vanish. On the right-hand side, $\partial_n\rho\big|_p \neq 0$, since the surface is defined by $\rho\to 0$, so nearby surfaces must have $\rho\ne0$.  Thus, the right-hand side vanishes when
\begin{equation}
\frac{1}{R_1}+\frac{1}{R_2}=k_1+k_2=0,
\end{equation}
where $k_1$ and $k_2$ are the principle curvatures at $p$, so the mean curvature at $p$ is zero.  There are several way to write curvatures on surfaces \cite{gray06}. We conform to the convention where the curvature is positive if the curve turns in the same direction as the surface normal ($C_2$), and negative otherwise ($C_1$).  Since $p$ is any point on the surface of constant vanishing $\rho$, the mean curvature vanishes everywhere on this surface, so the surface with $\rho=0$ is a minimal surface.

To show that surfaces of vanishing charge density are minimal, we have used the Vienna {\it ab initio} simulation package (VASP) \cite{kresse93,kresse96,kresse96b,kresse99} for materials with different electronic properties. The energy cut-off was set to 520 keV and the ${\bf k}$-spacing was set to $0.02\ $\AA$^{-1}$. The reciprocal lattice is sampled with a Monkhorst--Pack \cite{monkhorst76,monkhorst77} scheme for Na and a Gamma--centered scheme for Al and Cu. Exchange-correlation (XC) functionals and fast Fourier transform (FFT) mesh grids chosen for plotting surfaces of a given charge density $\rho$ are listed in Table.~\ref{table2}.  We use the local density approximation (LDA) \cite{kohn65} for Na, and the non-empirical generalized-gradient-approximation of Perdew, Burke and Ernzerhof (GGA-PBE) functional \cite{perdew96} for Al and Cu. Lattice parameters were obtained from experiments \cite{jain13} as the starting points for structural relaxation.

\begin{table}[t!]
\caption{\label{table2} The phases, exchange-correlation (XC) functionals, and the number of mesh grid points used for the calculations are reported here.}
\begin{ruledtabular}
\begin{tabular}{lccc}
Material & Phase & XC & FFT Mesh  \\
\hline
\hskip5pt Na & BCC & LDA &  $140\times 140 \times 140$ \\
\hskip5pt Cu & FCC & GGA-PBE & $100\times 100\times 100$ \\
\hskip5pt Al & FCC & GGA-PBE & $100\times 100 \times 100$ \\
\end{tabular}
\end{ruledtabular}
\end{table}

After optimizing the structural parameters of the crystal, we calculate the self-consistent charge density $\rho$ of the ground state from the Kohn--Sham orbitals \cite{martin20}. Our calculation is conducted over one unit cell, so we set $n_{\text{cell}}$ to one and the FFT grid density is the number of grid points that divides the unit cell in real space. As $\rho$ is a number density, we divide by the crystal volume to obtain the dimension $L^{-3}$.

We use the root-mean-square (RMS) to estimate the deviations of the surfaces of constant charge density $S_\rho$ from its corresponding TPMS, denoted by $S$.  To calculate the RMS, we take a grid of $100\times100$ mesh points on each elementary piece of the unit cell of the TPMS, from which the entire unit cell can be generated from the operations of the point group. Thus, there are $8\times10^4$ points on the P surface, $1.6\times10^5$ points on the I-WP surface, and $3.2\times10^5$ points on the F-RD surface. $S_\rho$ is formed by $n_{ab}$ points on the FFT mesh with their corresponding charge density $\rho$ falling in an interval $[\rho_a,\rho_b]$. Thus, $\rho\approx{1\over2}(\rho_a+\rho_b)$. We then scale the TPMS, calculate the smallest distance $d_i$, $i\in n_{ab}$, between each point on $S_\rho$ and $S$, and establish a data set of the smallest distances, which we use to obtain the RMS.  The RMS is calculated as: 
\begin{equation}
{\rm RMS}=\sqrt{\sum_i^{n_{ab}}d_i^2\over n_{ab}}.
\end{equation}

\begin{figure}[t!]
\centering
\includegraphics[width=1\linewidth]{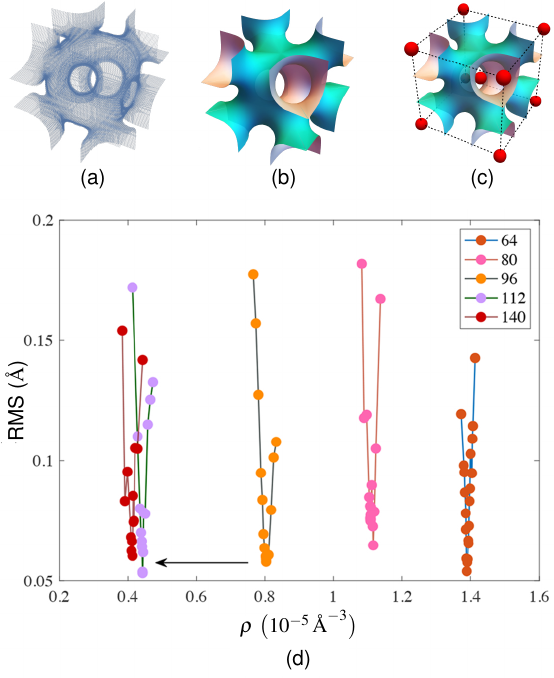}
\caption{(a) Charge densities of constant $\rho$ of Na in a BCC lattice. Plot of constant $\rho$ at given values against (b) the I-WP surface. (c) Plot of the BCC lattice against the I-WP TPMS. Red spheres in (c) represent atoms. (d) RMS against charge density $\rho$ of Na in the BCC lattice. The points are where calculations were carried out and the lines are guides for the eye. The arrow points to the minimum of the RMS and the mesh densities are shown in the legend.}
\label{fig5}
\end{figure}
\begin{figure}[t!]
\centering
\includegraphics[width=1\linewidth]{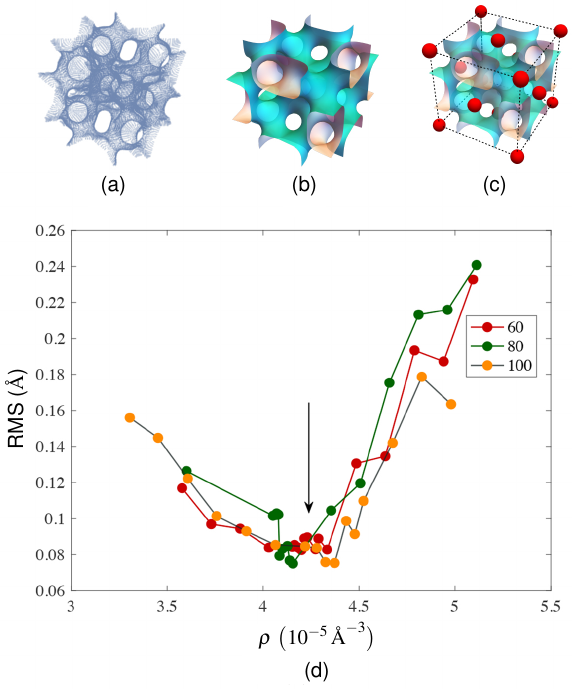}
\caption{(a) Charge densities of constant $\rho$ of Al in an FCC lattice. Plot of constant $\rho$ at given values against (b) the F-RD surface. (c) plots the FCC and BCC lattice against the F-RD surface. Red spheres in (c) represent atoms. (d) RMS against charge density $\rho$ of Al in the FCC lattice. The points are where calculations were carried out and the lines are guides for the eye. The arrow points to the minimum of the RMS and the mesh densities are shown in the legend.}
\label{fig6}
\end{figure}

\begin{figure}[t!]
\centering
\includegraphics[width=1\linewidth]{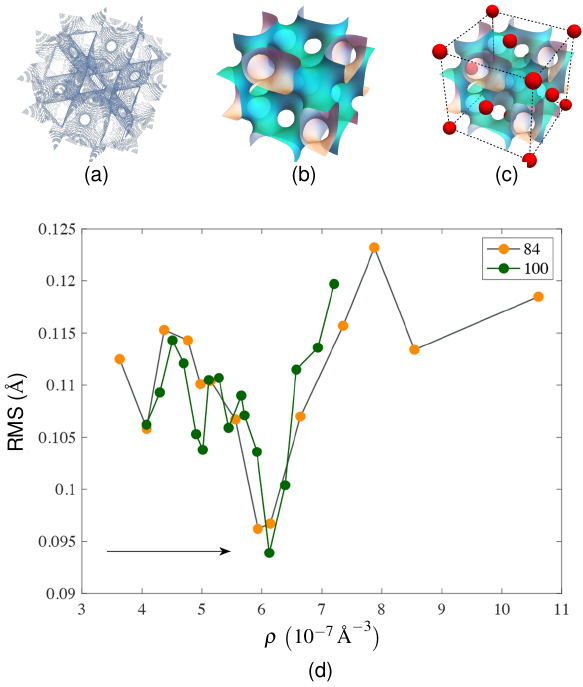}
\caption{(a) Charge densities for constant $\rho$ of Cu in an FCC lattice. (b) The F-RD surface. (c) plots the FCC lattice against the F-RD TPMS. Red spheres in (c) represent atoms. (d) RMS against charge density $\rho$ of Cu in an FCC lattice. The points are where calculations were carried out and the lines are guides for the eye. The arrow points to the minimum of the RMS and the mesh densities are shown in the legend.}
\label{fig7}
\end{figure}

Thus, the lower the dispersion of computed points from its TPMS, the lower the RMS, so the more $S_\rho$ resembles $S$.

Sodium (Na) is in the first column of the periodic table which, along the other alkali elements, has a single $s$ electron outside a closed shell. Sodium has atomic number 11, with the electronic configuration $1s^2\,2s^2\,2p^6\,3s^1$. The band structure has a single half-filled $3s$-band, with some $3p$ admixture for higher bands. The bands along the principal directions in ${\bf k}$-space are nearly identical, and surfaces of constant energy remain nearly spherical until they are close to the Brillouin zone boundary \cite{ham62}. Sodium has the BCC structure (space group $Im\overline{3}m$) at room temperature under atmospheric pressure. 

Figure~\ref{fig5}(d) shows the RMS against the charge density for Na in the BCC lattice for the indicated FFT mesh densities. The minimum decreases significantly with increasing mesh density and, for all meshes, the RMS shows sharp minima. Convergence with increasing mesh density is clearly evident. The converged minimum of $4.3\times10^{-6}\text{\AA}^{-3}$ is the best fit obtained. The converged surface is shown in Fig.~\ref{fig5}(a), together with the I-WP TPMS corresponding to the BCC lattice of Na in Fig~\ref{fig2}(b), and the positions of the Na atoms against the TPMS in Fig.~\ref{fig5}(c). Thus, for BCC Na with the LDA, we see clear convergence of the surfaces of constant charge density to the I-WP minimal surface. 

Aluminum (Al) has the FCC lattice (space group $Fm\overline{3}m$) at standard temperature and pressure. With atomic number 13, aluminum has three valence electrons per atom:~$3s^2\,3p$. The lower parabolic-shaped bands lie below the Fermi energy and are, therefore, completely filled and associated mainly with $\sigma$ bonds formed from two 3s electrons with some $3p$ admixture.  These lower bands strongly resemble bands arising free electrons \cite{segall61}. The third valence electron occupies the second and third $p$-bands, though with some $3s$ contribution as well. This band has $\pi$ character along the nearest-neighbor directions.  

Figure~\ref{fig6}(d) shows the RMS against the charge density for FCC Al as a function of the indicated mesh density. The arrow indicates the minimum value of the RMS. For all FFT meshes, the RMS shows sharp minima in the same region of charge density, with convergence with increasing mesh density and the minimum shifted slightly to lower charge densities, converging to $4.2\times10^{-5}$\AA$^{-3}$. The converged surface is shown in Fig.~\ref{fig6}(a), the F-RD TPMS corresponding to the FCC Al in Fig.~\ref{fig6}(b), and the positions of the Al atoms with respect to the TPMS are shown in Fig.~\ref{fig6}(c)]. Thus, for FCC Al with the GGA-PBE XC potential, surfaces of constant charge density converge to the F-RD minimal surface.

Copper (Cu) also has the FCC lattice (space group $Fm\overline{3}m$) at standard temperature and pressure. Copper has atomic number 29 with 11 valence electrons per atom:~$3d^{10}\,4s^1$.  The $4s$ electrons form wide bands in all high-symmetry directions, but when they cross the $3d$ bands, hybridization splits the $s$ and $d$ bands with the same symmetry near their initial crossing point. The remaining bands are unhybridized and form narrow bands \cite{segall62} across the Brillouin zone \cite{segall62}.

Figure~\ref{fig7}(d) shows the RMS against the charge density for Cu in the FCC lattice as a function of the indicated mesh densities. For all meshes, the RMS shows sharp minima in the same region of constant density, with convergence increasing slightly with the increasing mesh density. The best fit to the F-RD surface, such that the converged minimum appears, is around $6.1\times 10^{-7}\text{\AA}^{-3}$. The converged surface, the F-RD surface and the positions of the Cu atoms with respect to the TPMS are shown in Fig.~\ref{fig5}.Thus, for FCC Cu with the GGA-PBE XC potential, we see convergence of the surfaces of constant charge density to the F-RD minimal surface.

Our calculations provide fundamental physical realizations of minimal surfaces in terms of electronic characterizations of crystalline materials, rather than simply structural models.  Minimal surfaces have been used to describe many classes of materials \cite{hyde96}, but the first to recognize the importance of TPMS for solid state transitions were Hyde and Andersson \cite{hyde85,hyde86}. These authors associated TPMS with the initial and final states of a martensitic transformation. Although they were interested in iron, the same ideas can be applied to other martensitic materials and to the shape memory effect. We have discussed some of these ideas in \cite{yin24}, but a more complete discussion will be presented elsewhere.


\begin{thebibliography}{50}
\frenchspacing

\bibitem{hyde96} S. T. Hyde, Z. Blum, T. Landh, S. Lidin, B. W. Ninham, S. Andersson, and K. Larsson, {\it The Language of Shape:~The Role of Curvature in Condensed Matter Physics, Chemistry, and Biology} (Elsevier, Amsterdam, 1997).

\bibitem{fomenko91} A. T. Fomenko and A.A. Tuzhilin, {\it Elements of the Geometry and Topology of Minimal Surfaces in Three-Dimensional Space} (American, Mathematical Society, Providence, R. I., 1991).

\bibitem{carmo16} M. P. D. Carmo, {\it Differential Geometry of Curves and Surfaces} (Dover, New York, NY, 2016).

\bibitem{schoen70} A. H. Schoen, {\it Infinite periodic minimal surfaces without self-intersections}, Tech. Rep. TN D-5541 (NASA, Washington, D,. C., 1970).

\bibitem{schwarz90} H. A. Schwarz, {\it Gesammelte Mathematische Abhandlungen} (Springer, Berlin, 1890).

\bibitem{andersson88} S. Andersson, S. T. Hyde, K. Larsson, and S. Lidin, Minimal surfaces and structures: from inorganic and metal crystals to cell membranes and biopolymers, Chem. Rev. {\bf 88}, 221--242 (1988).\\
{\tt https://doi.org/10.1021/cr00083a011}

\bibitem{torquato04} S. Torquato and A. Donev, Minimal surfaces and multifunctionality, Proc. R. Soc. Lond. A {\bf 460}, 1849--1856 (2004).\\
{\tt https://doi.org/10.1098/rspa.2003.1269}

\bibitem{kapfer11} S. C. Kapfer, S. T. Hyde, K. Mecke, C. H. Arns, and G. E. Schr\"{o}der-Turk, Minimal surface scaffold designs for tissue engineering, Biomater. {\bf 32}, 6875--6882 (2011).\\
{\tt https://doi.org/10.1016/j.biomaterials.\\2011.06.012}

\bibitem{han18} L. Han and S. Che, An overview of materials with triply periodic minimal surfaces and related geometry: From biological structures to self-assembled systems, Adv. Mater. {\bf 30}, 1705708 (2018).\\
{\tt https://doi.org/10.1002/adma.201705708} 

\bibitem{alketan19} O. Al-Ketan and R. K. Al-Rub, Multifunctional mechanical metamaterials based on triply periodic minimal surface lattices, Adv. Eng. Mater. {\bf 21}, 1900524 (2019).\\
{\tt https://doi.org/10.1002/adem.201900524}

\bibitem{schnering87} H. G. von Schnering and R. Nesper, How nature adapts chemical structures to curved surfaces, Angew. Chem.  Int. Ed. {\bf 99}, 1059--1080 (1987).\\
{\tt https://doi.org/10.1002/anie.198710593}

\bibitem{schnering91} H. G. von Schnering and R. Nesper, Nodal surfaces of Fourier series: Fundamental invariants of structured matter, Z. Phys. B -- Condensed Matter {\bf 83}, 407--412 (1991).\\
{\tt https://doi.org/10.1007/BF01313411}

\bibitem{mackay93} A. L. Mackay, Crystallographic surfaces, Proc. R. Soc. Lond. A {\bf 442}, 47--59 (1993).\\
{\tt https://doi.org/10.1098/rspa.1993.0089}

\bibitem{gandy01} P. J. Gandy, S. Bardhan, A. L. Mackay, and J. Klinowski, Nodal surface approximations to the P, G, D and I-WP triply periodic minimal surfaces, Chem. Phys. Lett. {\bf 336}, 187--195 (2001).\\
{\tt https://doi.org/10.1016/S0009-2614(00)01418-4}

\bibitem{gandy02} P. J. F. Gandy and J. Klinowski, Geometric quantization of curvature energy in equipotential surfaces of ionic crystals, J. Chem. Phys. {\bf 116}, 9431--9434 (2002).\\
{\tt https://doi.org/10.1063/1.1471246}

\bibitem{landau81} L. D. Landau and E. M. Lifshitz, {\it Physical Kinetics}, Vol. 10 of {\it Course of Theoretical Physics} (Pergamon, Oxford, 1981).

\bibitem{jungel09} A. J\"{u}ngel, {\it Transport Equations for Semiconductors}, Lect. Notes Phys. 773 (Springer, Berlin, 2009).

\bibitem{reif65} F. Reif, {\it Fundamentals of Statistical and Thermal Physics} (MacGraw-Hill, New York, 1965).

\bibitem{gray06} E. Abbena, S. Salamon, and A. Gray, (2006). {\it Modern Differential Geometry of Curves and Surfaces with Mathematica} 3rd ed. (Chapman and Hall/CRC, Boca Raton, FL, 2006).

\bibitem{kresse93} G. Kresse and J. Hafner, {\it Ab initio} molecular dynamics for liquid metals, Phys. Rev. B {\bf 47}, 558--561 (1993).\\
{\tt https://doi.org/10.1103/PhysRevB.47.558}

\bibitem{kresse96} G. Kresse and J. Furthm\"{u}ller, Efficient iterative schemes for {\it ab initio} total-energy calculations using a plane-wave basis set, Phys. Rev. B {\bf 54}, 11169--11186 (1996).\\
{\tt https://doi.org/10.1103/PhysRevB.54.11169}

\bibitem{kresse96b} G. Kresse and J. Furthm\"{u}ller, Efficiency of ab-initio total energy calculations for metals and semiconductors using a plane-wave basis set, Comput. Mat. Sci. {\bf 6}, 15--50 (1996).\\
{\tt https://doi.org/10.1016/0927-0256(96)00008-0}

\bibitem{kresse99} G. Kresse and D. Joubert, From ultrasoft pseudopotentials to the projector augmented-wave method, Phys. Rev. B {\bf 59}, 1758--1775 (1999).\\
{\tt https://doi.org/10.1103/PhysRevB.59.1758}

\bibitem{monkhorst76} H. J. Monkhorst and J. D. Pack, Special points for Brillouin-zone integrations, Phys. Rev. B {\bf 13}, 5188--5192 (1976).\\
{\tt https://doi.org/10.1103/PhysRevB.13.5188}

\bibitem{monkhorst77} H. J. Monkhorst and J. D. Pack, ``Special points for Brillouin-zone integrations'' -- a reply, Phys. Rev. B {\bf 16}, 1748--1749 (1977).\\
{\tt https://doi.org/10.1103/PhysRevB.16.1748}

\bibitem{kohn65} W. Kohn and L. J. Sham, Self-consistent equations including exchange and correlation effects, Phys. Rev. {\bf 140}, A1133--A1138 (1965).\\
{\tt https://doi.org/10.1103/PhysRev.140.A1133}

\bibitem{perdew96} J. P. Perdew, K. Burke, and M. Ernzerhof, Generalized gradient approximation made simple, Phys. Rev. Lett. {\bf 77}, 3865--3868 (1996).\\
{\tt https://doi.org/10.1103/PhysRevLett.77.3865}

\bibitem{jain13} A. Jain, S. P. Ong, G.Hautier, W. Chen, W. D. Richards, S. Dacek, S. Cholia, D. Gunter, D. Skinner, G. Ceder, and K. A. Persson, Commentary: The Materials Project: A materials genome approach to accelerating materials innovation, APL Mater. {\bf 1}, 011002 (2013).\\
{\tt https://doi.org/10.1063/1.4812323}

\bibitem{martin20} R. M. Martin, {\it Electronic Structure: Basic Theory and Practical Methods} 2nd edn (Cambridge University Press, Cambridge, 2020).

\bibitem{ham62} F. S. Ham, Energy bands of alkali metals. I. Calculated bands, Phys. Rev. {\bf 128}, 82--97 (1962).\\
{\tt https://doi.org/10.1103/PhysRev.128.82}

\bibitem{segall61} B. Segall, Energy bands of aluminum, Phys. Rev. {\bf 124}, 1797--1806 (1961).\\
{\tt https://doi.org/10.1103/PhysRev.124.1797}

\bibitem{segall62} B. Segall, Fermi surface and energy bands of copper, Phys. Rev. {\bf 125}, 109--122 (1962).\\
{\tt https://doi.org/10.1103/PhysRev.125.109}

\bibitem{klinowski96} J. Klinowski and A. L. Mackay and H. Terrones,  Curved surfaces in chemical structure, Phil. Trans. R. Soc. Lond. {\bf A354}, 1975--1987 (1996).\\
{\tt https://doi.org/10.1098/rsta.1996.0086}

\bibitem{mackay94} A. L. Mackay, Periodic minimal surfaces from finite element methods, Chem. Phys. Lett. {\bf 221}, 317--321 (1994).\\
{\tt https://doi.org/10.1016/0009-2614(94)00256-8}

\bibitem{hyde85} S. T. Hyde and S. Andersson, Differential geometry of crystal structure descriptions, relationships and phase transformation, Z. Kristallogr. Cryst. Mater. {\bf 170}, 225--239  (1985) .\\
{\tt https://doi.org/10.1524/zkri.1985.170.14.225} 

\bibitem{hyde86} S. T. Hyde and S. Andersson, The martensite transition and differential geometry, Z. Kristallogr. Cryst. Mater. {\bf 174}, 225--236  (1986).\\
{\tt https://doi.org/10.1524/zkri.1986.174.14.225}  

\bibitem{yin24} M. Yin and D. D. Vvedensky, Shape memory effect and the topology of minimal surfaces, Symmetry {\bf 16}, 1187 (2024).\\
{\tt https://doi.org/10.3390/sym16091187}

\end{thebibliography}
\end{document}